\documentclass[reprint,amsmath,amssymb,aps,prb, superscriptaddress, showkeys]{revtex4-2}

\usepackage{graphicx}
\usepackage{float}
\usepackage{dcolumn}
\usepackage{xcolor}
\usepackage{soul}
\usepackage[version=4]{mhchem}
\usepackage{layouts}
\usepackage{booktabs}
\usepackage{siunitx}

\newcolumntype{d}[1]{D{.}{.}{#1}}

\begin{document}

\title{Toward the Ultimate Limit: Elemental Metals in One Dimension}

\author{Mohammad Bagheri}
\affiliation{Nanoscience Center, Department of Physics, University of Jyv\"askyl\"a, 40014 Jyv\"askyl\"a, Finland}
\author{Kameyab Raza Abidi}
\affiliation{Nanoscience Center, Department of Physics, University of Jyv\"askyl\"a, 40014 Jyv\"askyl\"a, Finland}
\author{Sushree Sarita Sahoo}
\affiliation{Nanoscience Center, Department of Physics, University of Jyv\"askyl\"a, 40014 Jyv\"askyl\"a, Finland}
\author{Sukhbir Singh}
\affiliation{Department of Physics, Rayat Bahra University, Chandigarh‑Ropar NH 205, Greater Mohali, Punjab 140103, India}
\author{Pekka Koskinen}
\email{pekka.j.koskinen@jyu.fi}
\affiliation{Nanoscience Center, Department of Physics, University of Jyv\"askyl\"a, 40014 Jyv\"askyl\"a, Finland}
\date{\today}

\begin{abstract}
Low-dimensional materials exhibit extraordinary properties that make them promising candidates for advanced technologies. Although they have been investigated extensively, most of the research has focused on layered two-dimensional (2D) materials. Here, inspired by recent advances in atomically thin metallenes, we further reduce dimensionality and use density-functional theory simulations to study the geometry, energetics, elasticity, and electronic structure of $40$ non-magnetic one-dimensional (1D) atomic chains of elemental metals. We find that nearly all chains have a buckled ground state, nine chains are distorted, and three chains---Cd, Hg, and Sr---are semiconducting with an electronic gap. We also find that transition metals retain a substantial fraction of their 3D bulk cohesive energy even in 1D chains.
We assessed chains' dynamical stabilities by molecular dynamics simulations and found that $26$ of them are thermodynamically stable at $100$~K. Finally, we performed chain pulling simulations to investigate the straightening dynamics of selected stable chains.
Given that experimental techniques have recently reached the 1D-chain limit, our systematic study provides a foundation and timely guide to accelerating synthesis and characterization of these materials. 

\end{abstract}

\maketitle

\section{Introduction}
Low-dimensional materials continue to attract interest because of the novel phenomena associated with reduced dimensionality. Low dimensionality reorganizes bonding and symmetry in ways that enable new functionalities across optics \cite{Du21_NComm, Wei21_AM, Turunen22_NRevPhys}, magnetism \cite{Gibertini2019, Klein23_ACSNano}, electronics \cite{Radisavljevic2011, Manzeli2017, Xiang20_Sci, Han19_NComm}, sensing \cite{Kim_2018, mx_sens, Zhou_2021}, catalysis \cite{Deng2016, D4NR00147H, Zhang_2025}, and energy technologies \cite{TAO2020, Qi2024}.
New physics has driven research to focus primarily on two-dimensional (2D) materials, where van der Waals (vdW) exfoliation provides a natural route to reduced dimensions. In the meantime, other low-dimensional materials, such as one-dimensional (1D) metals, have remained largely overlooked.

Some interest has been directed toward quasi-one-dimensional chains, but their studies have been confined to specialized substrates \cite{SCHMIDT_1997, 1D_gold_2004, Gambardella_2002, MA_2022, BALANDIN_2022}. Early efforts utilized angle-resolved photoemission spectroscopy (ARPES) and scanning tunneling microscopy (STM) to probe the electronic structure of gold chains on silicon surfaces \cite{Crain_2004}, revealing a clear signature of broken symmetry \cite{Erwin_2013} and a structural transition in atomic chains \cite{Polei_2013}. 
More recently, physical and electroninc properties of Pb and Bi chains on surfaces such as Si \cite{Mihalyuk_2020}, Ag \cite{Sheverdyaeva_2022} and InAs \cite{Mihalyuk_2024, Sheverdyaeva_2024} were investigated. However, supported 1D chains become electronically coupled to their substrates, which masks the chains' \emph{intrinsic} properties and limits the exploration of the pure 1D limit \cite{PFNUR_2024, Sangmo_2015}.

Although 1D materials have been observed and investigated before \cite{Yanson1998}, their landscape has been overturned by recent breakthroughs in synthesis. Modern exfoliation and electron-beam processing methods have finally enabled researchers to approach the physical 1D limit, achieving suspended individual atomic chains \cite{Teeter_2024}. 
Additionally, the emergence of advanced experimental exfoliation methods that break covalent or ionic bonds \cite{Guan_2017, Anasori2017, Balan_2018, PuthirathBalan2018, Jiang2022, Kaur_2022}, rather than the traditional mechanical exfoliation of layered vdW crystals, has made the formation of 1D chains more feasible.
These experimental strides create an urgent need for computational explorations that can identify and characterize 1D candidates.
On the other hand, while the exotic properties of 1D materials are being widely mapped through high-throughput screening \cite{Hadeel_2023, Khazaei_2023} and materials databases \cite{FCDimen, Yanbing_2023, MLFCDimen}, computational studies on 2D metallenes have largely focused on stability frameworks \cite{atlas_2018, gentle} and interfacial engineering \cite{Edge_2025, Abidi_2025, interface}. Consequently, research focused specifically on 1D elemental metals remains notably lacking.

Therefore, in this article, we employed density-functional theory (DFT) simulations to study the intrinsic properties of 40 nonmagnetic 1D chains of elemental metals. For each chain, we identified ground-state geometry, bond length, elastic modulus $C_{11}$ (1D Young modulus), and electronic band structure. Among the various structural and energetic trends of all metals, we found three perfectly linear and nine distorted ground states, and an opening of the band gap in Cd, Hg, and Sr chains.
We also constructed longer chains and performed molecular dynamics (MD) simulations to evaluate their dynamical stabilities. We found that $26$ out of $40$ chains remain stable at $100$~K. Finally, we performed pulling MD simulations of Au and Na chains and obtained insights into their straightening dynamics.

\begin{figure*}
\centering
  \includegraphics[width=\textwidth]{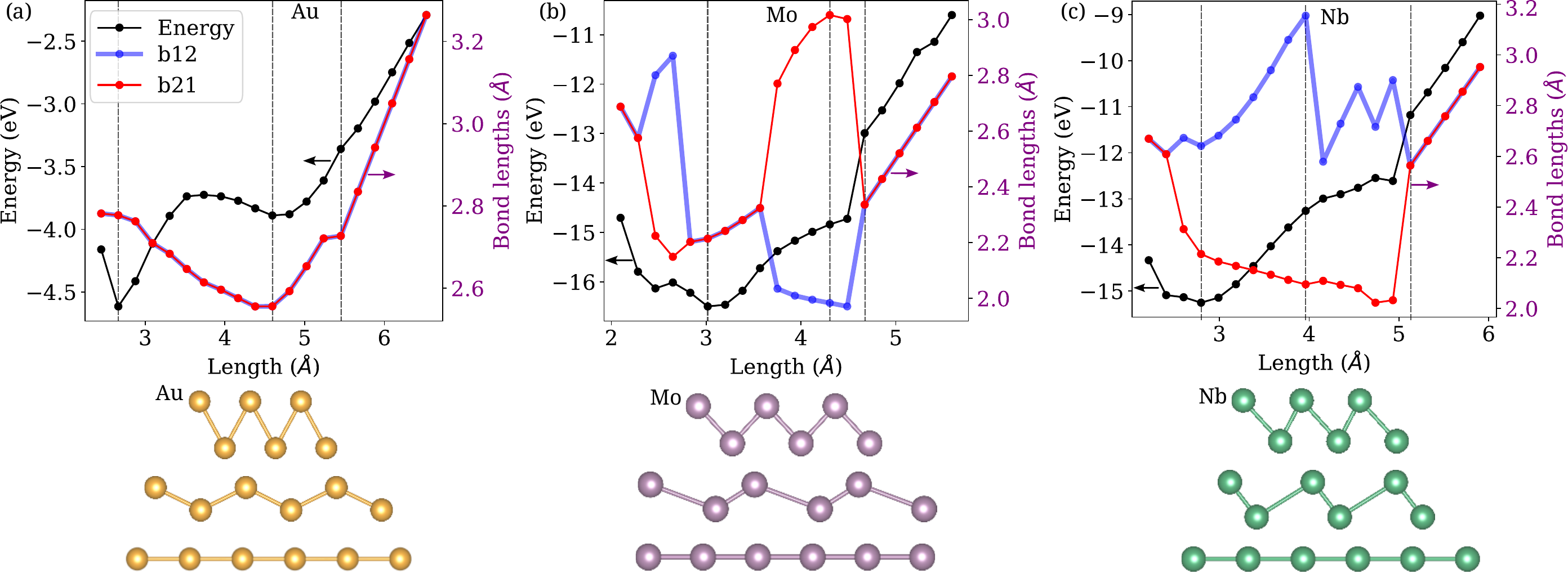}
\caption{\label{fig:chains}
 Energies and bond lengths as a function of two-atom cell length during scanning cell lengths from -10\% to 20\% of initial bond lengths. Schematic structures of chains at specific cell lengths are shown below (upper at shorter, lower at larger cell lengths). The corresponding cell lengths of each chemical structure are indicated with vertical dashed lines in the top panels.
}
\end{figure*}

\section{Methods and materials}

\subsection{Computational Methods}

The density-functional theory (DFT) calculations were performed in the plane-wave (PW) mode with $800$~eV cutoff energy \cite{pw} of the GPAW code \cite{gpaw1, gpaw3, ase-paper} using the Perdew-Burke-Ernzerhof (PBE) exchange and correlation (xc) functional \cite{pbe, GGA, libxc}. Long-range effects and correcting for the weak vdW interactions were accounted for by the D3 correction of GPAW, which uses Stefan Grimme’s DFT-D3 code \cite{d3-1, d3-2, d3-3}. The calculations were spin-polarized and converged with respect to the vacuum, the ${\bf k}$-point grids, and the plane-wave cutoff. All calculations used a two-atom unit cell with adjustable cell length in the periodic direction and $20$-\AA\ vacuum regions in non-periodic directions. The ${\bf k}$-point sampling was Monkhorst-Pack \cite{PhysRevB.13.5188, PhysRevB.16.1748} with a default $1 \times 1 \times 15$ mesh, and a denser $1 \times 1 \times 20$ mesh for some heavier elements to ensure convergence. For reference calculations of three-dimensional (3D) bulk, we used $12 \times 12 \times 12$ mesh. Under these settings, all structures were relaxed using the BFGS algorithm to forces below $1$~meV/\AA. The atomic energies were obtained from $\Gamma$-point calculations of single atoms in $20$~\AA\ cubic cells.

Electronic band structures were calculated along the periodic path ($\Gamma$--Z) using the PBE xc-functional. The electron density was determined self-consistently on a uniform $1 \times 1 \times 30$ k-point grid. From this density, the PBE band structure was computed non-self-consistently at $201$ k-points distributed along the band path.  
For semiconducting systems, band structures were calculated using HSE06 \cite{HSE06} on top of a PBE calculation using the converged PBE ground-state density and $401$ k-points distributed along the band path.
The density of states (DOS) and projected density of states (PDOS) were calculated using linear tetrahedron interpolation of energy eigenvalues obtained from a ground-state calculation.

Molecular dynamics (MD) simulations were performed for a supercell 10 times the unit cell. The $10$-ps MD runs used a time step of $2.5$ fs and a $100$ K Langevin thermostat with a friction coefficient of $0.002$~fs$^{-1}$.

\subsection{1D Chain Simulations}

We constructed the cell with a two-atom basis and an initial bond length ($d$), defined either as $2 \times$covalent atomic radii or the predicted lattice constants of the pristine 2D (hexagonal, square, and honeycomb) metallenes (Ref.~\citenum{atlas_2018}), whichever was shorter. We then scanned cell lengths $l$ from -10\% to 20\% and optimized structures starting from slightly asymmetrically perturbed initial guesses to prevent the geometry optimization from getting trapped in the high-symmetry saddle point of a perfectly linear chain. For cases in which energy convergence was not reached within the initial scanning range, we expanded the range. The ground-state structure was assigned as the lowest-energy chain in our scanning calculations. In addition to cell length $l$, the chain geometries were examined through bond lengths $b_{12}=\sqrt{(x_2 - x_1)^2+(z_2 - z_1)^2}$ and $b_{21}=\sqrt{(x_2 - x_1)^2+(l + z_2 - z_1)^2}$, whose ratio serves as an order parameter for distortion.

We performed these calculations for 40 metallic elements, excluding magnetic ones (Cr, Mn, Fe, Co, and Ni), whose magnetic complexity falls beyond the scope of this work.
It is worth noting that although magnetism is often observed in reduced dimensions, such as 2D metallenes \cite{Abidi_2025, D1TC00438G}, in 1D it did not appear; all metals remained non-magnetic, just like their 3D counterparts.

\section{Results}

\subsection{Ground-state geometry}

We begin by identifying the preferred ground-state geometries by analyzing energies and bond lengths as functions of the cell length. Physically, a perfectly linear metallic chain is often susceptible to electronic instabilities, and chains lower their energy by buckling or Peierls distortion, meaning dimerization with alternating bond lengths~\cite{peierls1955quantum}. Buckling allows orbital overlap in a more energetically favorable way than collinear bonding. The results for all elements are shown in the Supporting Information (Figs.~S1--S3); here we discuss a few illustrative examples.

For the gold chain, bond lengths are equal (b$_{12}$=b$_{21}$), and the ground state is prominently buckled (Fig.~\ref{fig:chains}a). However, many transition metals with partially filled $d$-bands become unstable against distortion. Such distortions occurred in Ti, V, Nb, Mo, W, Os, Ir, Pt, and Bi. 
For example, Mo and Nb chains have pronounced differences between b$_{12}$ and b$_{21}$. The ground-state of Mo is a symmetric buckled chain with equal bond lengths b$_{12}$=b$_{21}$=$2.21$~\AA\ (Fig.~\ref{fig:chains}b). However, away from the ground state, around $l\approx 4.5$~\AA, Mo exhibits a pronounced bond asymmetry (b$_{12}$=$1.98$~\AA~and b$_{21}$=$3.01$~\AA), indicative of a strong tendency toward Peierls dimerization where electron localization in short bonds may open a gap. The ground state of Nb, on the other hand, is itself greatly distorted (b$_{12}$=$2.64$~\AA\ and b$_{21}$=$2.21$~\AA). Indeed, Nb chain is severely distorted within the major part of the entire scanning range, distortions reaching up to b$_{12}$=$3.16$~\AA, b$_{21}$=$2.09$~\AA\ at $l=3.96$~\AA\ (Fig.~\ref{fig:chains}c). Yet another interesting case is the Os chain, in which the entire ground state region is strongly distorted. The bond-length disparity is large (b$_{12}$=$3.29$~\AA, b$_{21}$=$2.18$~\AA) and becomes even more pronounced in the most distorted configuration (b$_{12}$=$3.41$~\AA, b$_{21}$=$2.16$~\AA) (Fig.~S4). 

The perfectly linear chain is a ground state only for Re, Y, Sr, and Zn. For these elements, the electronic energy gain from distortion or buckling is insufficient to overcome the elastic energy of lattice deformation. However, the chain becomes linear for all elements beyond a certain cell length (Fig.~\ref{fig:chains}). On average, this onset length corresponds to $3.6 \pm 0.75 \times$the covalent atomic radii. In other words, buckling of 1D chains occurs at bond lengths around $1.8\times$the covalent atomic radii.

\subsection{Energy and bond lengths}

After identifying the ground-state geometries, we determined the energies and bond lengths of the optimized chains. The cohesive energy of the structure is defined as
\begin{equation}
    E_{coh} = E_{atom} - E/N,
\end{equation}
where $E_{atom}$ is the energy of a free atom, $E$ is the relaxed energy of the structure, and $N$ is the number of atoms in the unit cell.

\begin{figure}
\centering
  \includegraphics[width=\linewidth]{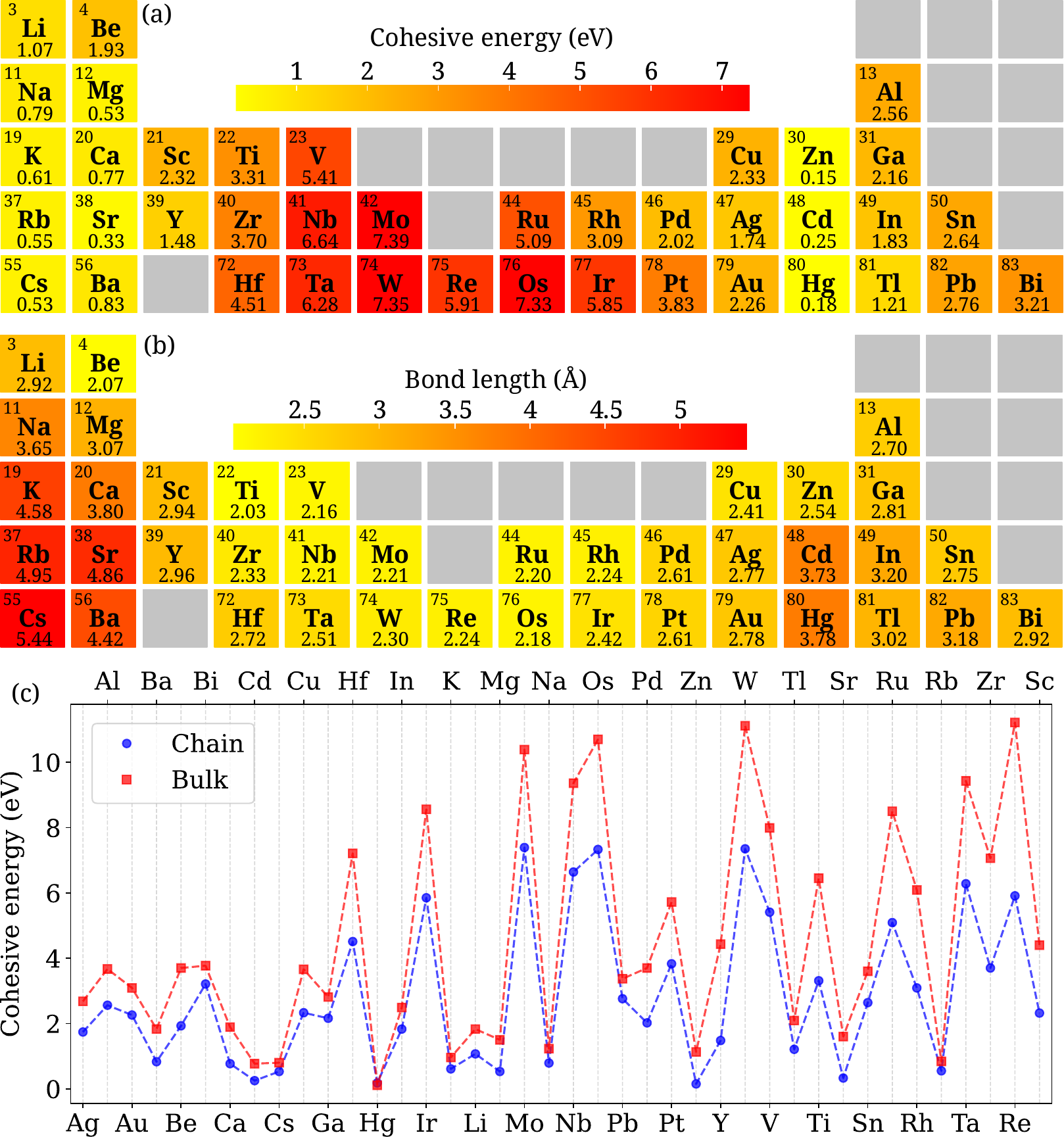}
\caption{\label{fig:Emin_b12}
 Energetic and geometric trends. (a) Cohesive energies and (b) bond lengths of the 1D chains in the ground-state. (c) Comparing cohesive energies of 1D chains to those of 3D bulks.
}
\end{figure}

The cohesive energies follow a parabolic trend across the transition metal series, driven by the gradual filling of the bonding and anti-bonding d-states. Consequently, the transition metals exhibit the strongest cohesion and the shortest bond lengths (Figs.~\ref{fig:Emin_b12}a,b). Specifically, the Group 6 elements (Mo, W) lie at the peak of this curve where the bonding states are maximally filled; Mo displays the highest cohesive energy ($7.39$~eV), retaining $71\%$ of its 3D bulk cohesion. This retention rate is characteristic of transition metals (typically $50$--$70\%$), where the strengthening of the remaining bonds compensates for the reduced coordination, causing the cohesive energy to scale sub-linearly with coordination number \cite{abidi2022optimizing}.
Notably, heavy p-block elements like Bi and Pb retain an even higher fraction of their bulk energy ($>80\%$), suggesting their directional p-orbitals adapt exceptionally well to one-dimensional geometries.

In contrast, elements with closed-shell configurations exhibit a disproportionate loss of cohesion in 1D. Zn possesses the lowest cohesive energy ($0.15$~eV), retaining only $13\%$ of its bulk value, while Sr retains only $21\%$. This considerable reduction occurs because the isotropic s-bonding in these groups relies heavily on high coordination numbers (3D packing) to force orbital overlap; in a 1D chain, this mechanism fails, leaving the atoms weakly bound.

Despite being a closed-shell element, Hg presents an anomaly to this trend and has cohesive energy ($0.18$~eV) that effectively exceeds its bulk value ($\sim 164\%$ retention) (Fig.~\ref{fig:Emin_b12}c). This behavior is likely driven by significant relativistic contraction and strong electron correlations, which can stabilize low-dimensional geometries in ways standard DFT functionals struggle to capture in the bulk limit~\cite{hg_2006, atlas_2018}.

Structurally, the bond lengths correlate inversely with bond strength. Ti possesses the shortest bond length ($2.03$~\AA) due to strong d-orbital overlap, while Cs exhibits the longest ($5.44$~\AA) due to weak s-orbital overlap. On average, ground state bond lengths correspond to $1.92 \pm 0.38 \times$the covalent atomic radii.
Notably, the bond lengths in 1D are systematically shorter than those in 2D hexagonal lattices or bulk phases. This contraction is a physical response to the lower coordination number as atoms are drawn closer to maximize the electron density and strength of the remaining bonds, a behavior analogous to that observed in 2D metallenes \cite{atlas_2018}.

\begin{figure}
\centering
  \includegraphics[width=\linewidth]{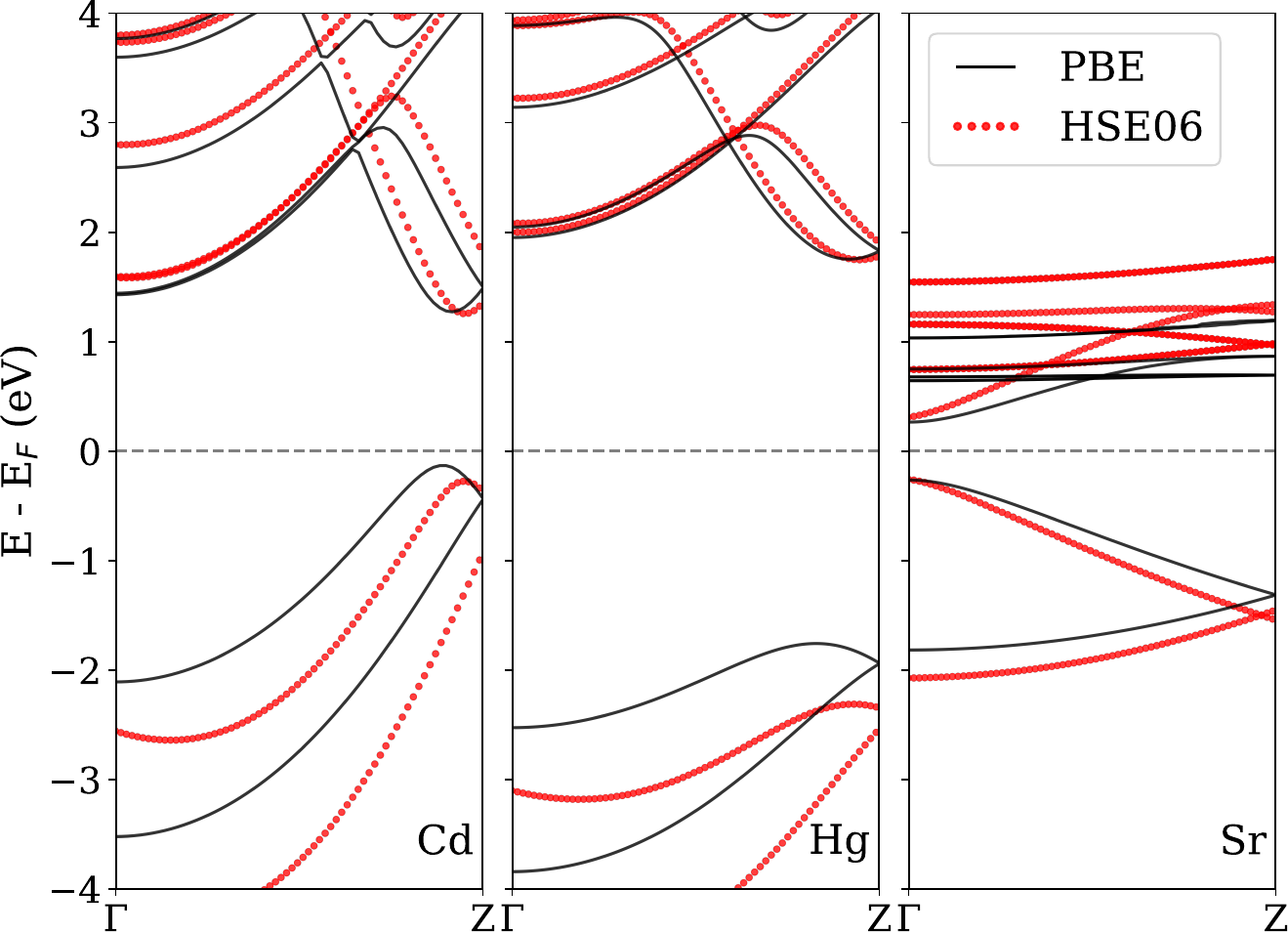}
\caption{\label{fig:bands}
PBE and HSE06 band structures for 1D chains with an electronic bandgap (Cd, Hg, and Sr).
}
\end{figure}

\begin{figure}
\centering
    \includegraphics[width=\linewidth]{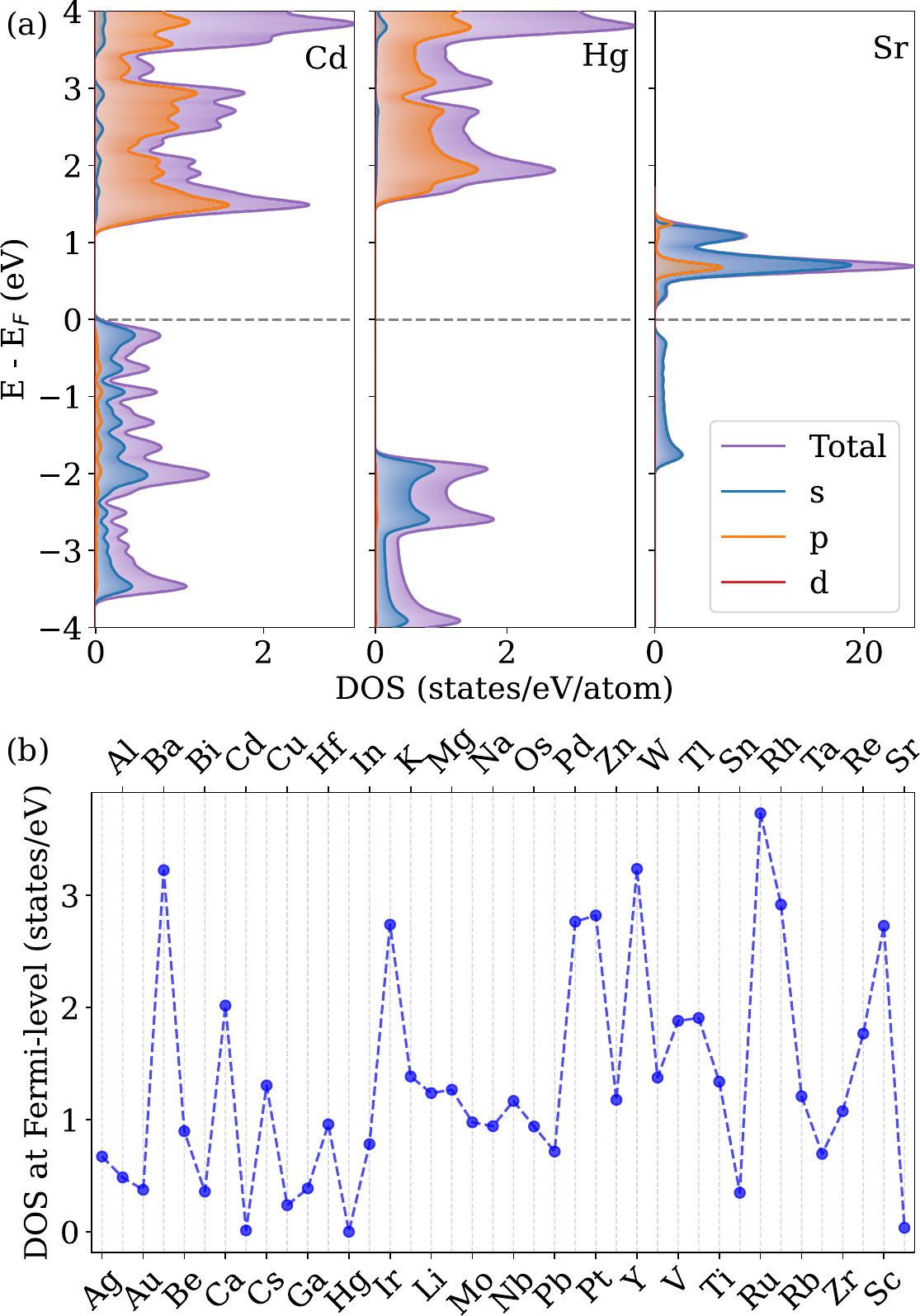}
\caption{\label{fig:dos}
(a) Projected density of states (PDOS) for 1D chains with an electronic bandgap (Cd, Hg, and Sr). (b) Density of states (DOS) of 1D chains at the Fermi-level.
}
\end{figure}

\subsection{Elasticity}
Energy curves also provide information about the elastic properties of the 1D chains. Under uniaxial deformation, the strain is $\varepsilon=(l_z-l_{z,0})/l_{z,0}$, where and $l_z$ is the strained and $l_{z,0}$ the unstrained length in $z$-direction. We define the elastic energy density per unit length as
\begin{equation}
   \Delta U(\varepsilon)=[E(\varepsilon)-E_0]/l_{z,0}=\tfrac12 C_{11}\varepsilon^2, 
\end{equation}
where $U$ is the energy density per unit length, $E$ is the energy, and $C_{11}$ is the elastic constant, equivalent to 1D Young's modulus.

The 1D elastic constant was thus obtained from the curvature of the total energy with respect to strain. The resulting moduli vary considerably in magnitude (Table~\ref{tab:c11}). The largest moduli are found for Re ($59.33$ eV/$\si{\angstrom}$), Ta ($31.16$ eV/$\si{\angstrom}$), and Os ($26.26$ eV/$\si{\angstrom}$), in agreement with their large cohesive energies. This trend is opposed by Nb, Mo, and W, which remain unexpectedly compliant. In general, alkali metals have small moduli, with the smallest of them for Cs ($0.67$~eV/$\si{\angstrom}$).

\begin{table}
\centering
\caption{Elastic constant $C_{11}$ (eV/$\si{\angstrom}$) for 1D chains.}
\label{tab:c11}
\begin{tabular}{c d{2} c d{2} c d{2}}
\toprule
Element & \multicolumn{1}{c}{$C_{11}$} & Element & \multicolumn{1}{c}{$C_{11}$} & Element & \multicolumn{1}{c}{$C_{11}$} \\ 
\midrule
Ag	&	1.63	&	Ir	&	13.47	&	Ru	&	18.27	\\
Al	&	2.71	&	K	&	1.29	&	Sc	&	11.07	\\
Au	&	6.63	&	Li	&	3.03	&	Sn	&	8.03	\\
Ba	&	0.89	&	Mg	&	1.32	&	Sr	&	2.98	\\
Be	&	3.66	&	Mo	&	10.19	&	Ta	&	31.16	\\
Bi	&	2.61	&	Na	&	1.17	&	Ti	&	19.80	\\
Ca	&	4.47	&	Nb	&	8.62	&	Tl	&	4.47	\\
Cd	&	1.23	&	Os	&	26.26	&	V	&	17.90	\\
Cs	&	0.67	&	Pb	&	13.85	&	W	&	25.33	\\
Cu	&	2.86	&	Pd	&	10.00	&	Y	&	8.01	\\
Ga	&	2.22	&	Pt	&	10.78	&	Zn	&	6.73	\\
Hf	&	7.05	&	Rb	&	1.85	&	Zr	&	17.49	\\
Hg	&	1.10	&	Re	&	59.33	&	~ & \multicolumn{1}{c}{~}	\\		
In	&	1.67	&	Rh	&	9.14	&	~ & \multicolumn{1}{c}{~}	\\		
\bottomrule
\end{tabular}
\end{table}

\subsection{Electronic structures}

To explore the electronic properties of the chains, we calculated their band structures and projected densities of states (PDOS).
Band structure calculations using the PBE functional revealed the opening of band gaps in three chains.
Among transition metals, we found a gap for Hg ($3.51$~eV) and Cd ($1.41$), and among alkaline-earth metals, we found a gap for Sr ($0.53$~eV) (Fig.~\ref{fig:bands}), indicating that the s-p band overlap present in the bulk is lifted due to the strong quantum confinement.

The band structures revealed that for Cd and Hg, both the conduction band minimum and the valence band maximum were located near the Z point. In contrast, for Sr it occurred near the $\Gamma$-point.
In all three cases, the alignment of band edges indicates a direct band gap, which is favorable for optical applications.
Since PBE often underestimates band gaps due to self-interaction errors, we recalculated the band structures using the HSE06 xc functional. With HSE06, the gaps increased to $4.05$~eV for Hg, $1.53$~eV for Cd, and $0.56$~eV for Sr (Fig.~\ref{fig:bands}). Note that, depending on the lattice, 2D Hg gave a gap of $1.55$--$3.55$~eV with DFT-HSE as well~\cite{fermi_met}.

We have also illustrated band structures of representative metallic chains, including Au, Be, Ga, Na and Nb (Fig.~S5). In contrast to the chains with band gap, the band structures of the metallic chains show robust, steep band crossings at $E_\mathrm{F}$.

The PDOS indicates that the s- and p-orbitals make the largest contributions near the Fermi level in all three semiconducting chains (Fig.~\ref{fig:dos}a), demonstrating that s-p hybridization dictates the gap formation. Furthermore, the DOS has sharp peaks at the band edges. These are characteristic van Hove singularities expected in 1D systems, indicating a high density of states available for optical transitions.
To compare the electronic structures between different chains and elements, we calculated the DOS at the Fermi energy (Fig.~\ref{fig:dos}b). All the metallic chains reflect the availability of states for electric conduction by a largely varying, non-zero DOS without an apparent general trend. In semiconductor chains, the DOS at the Fermi level for Hg, Sr, and Cd is practically zero.

\subsection{Thermodynamical stability of extended chains}

To ensure the thermodynamic stability of the chains, we examined them under more practical conditions. We constructed chains 10 times the ground-state unit-cell length and performed thermostatted molecular dynamics simulations.

\begin{figure}
\centering
  \includegraphics[width=\linewidth]{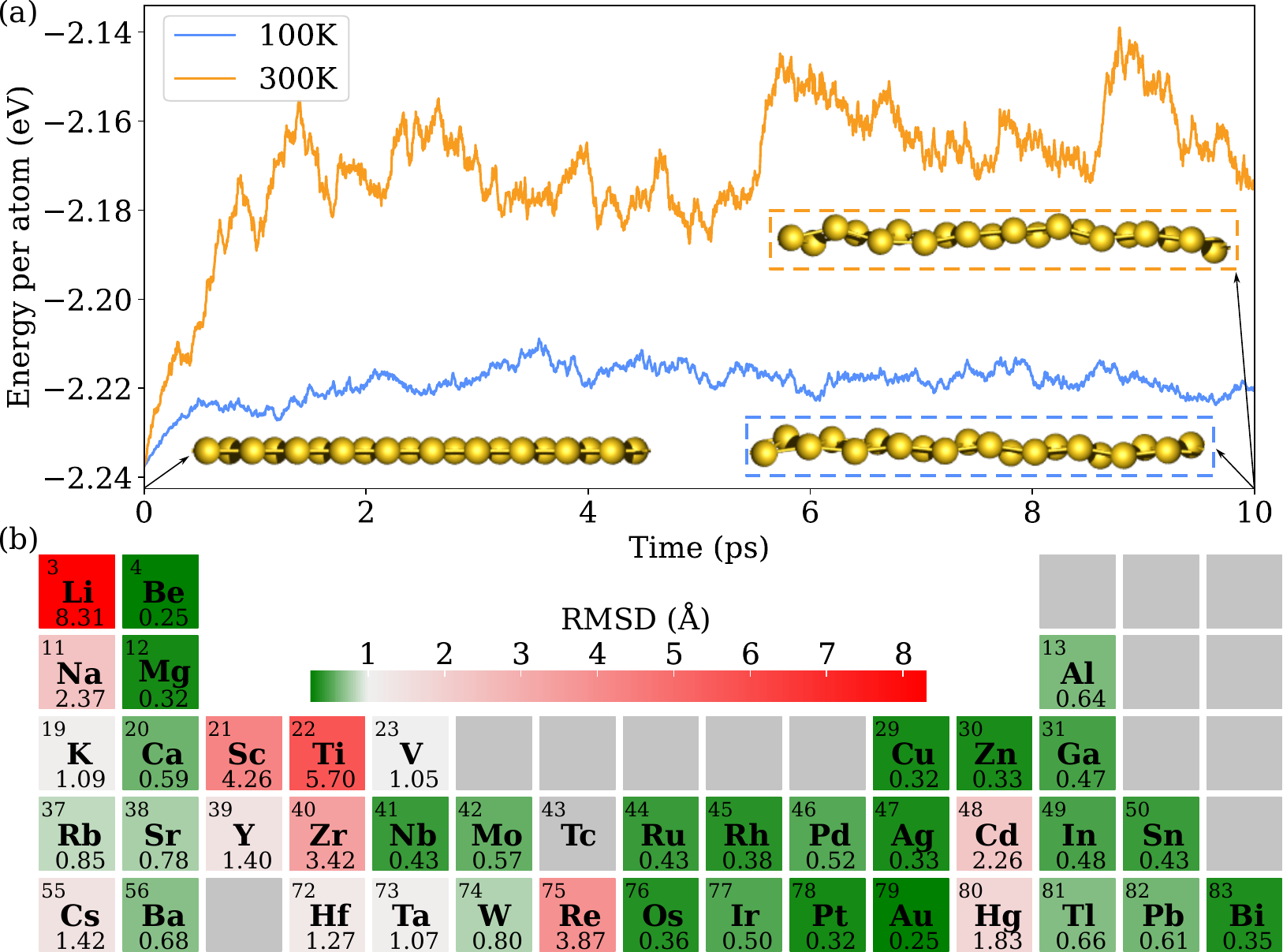}
\caption{\label{fig:md_rmsd}
 (a) Molecular dynamics of gold chain at two different temperatures, $100$ and $300$~K. The insets show the side views of the initial (0 ps) and final (10 ps) structures. (b) Heat maps of the RMSD from $100$-K MD simulations for all $40$ metals. Green color (RMSD$\lesssim 1$~\AA) corresponds to stable chains, red color (RMSD$\gtrsim 1$~\AA) to unstable chains.}
\end{figure}

Starting with gold chains at $300$~K and $100$~K, we observed that although gold chains can be semi-stable at room temperature, they are structurally constantly on the verge of breaking up. This tendency is reflected in the behavior of potential energy with sharp fluctuations (Fig.~\ref{fig:md_rmsd}a). Because gold is among the most stable metals in low dimensions, it is unlikely that chains of other elements would remain stable at room temperature. Upon lowering the temperature to $100$~K, the gold chain became clearly stable. Therefore, we performed all our MD simulations at $100$~K, which is cryogenic but still sufficiently large to characterize thermodynamical stability. 

\begin{figure*}
\centering
  \includegraphics[width=\textwidth]{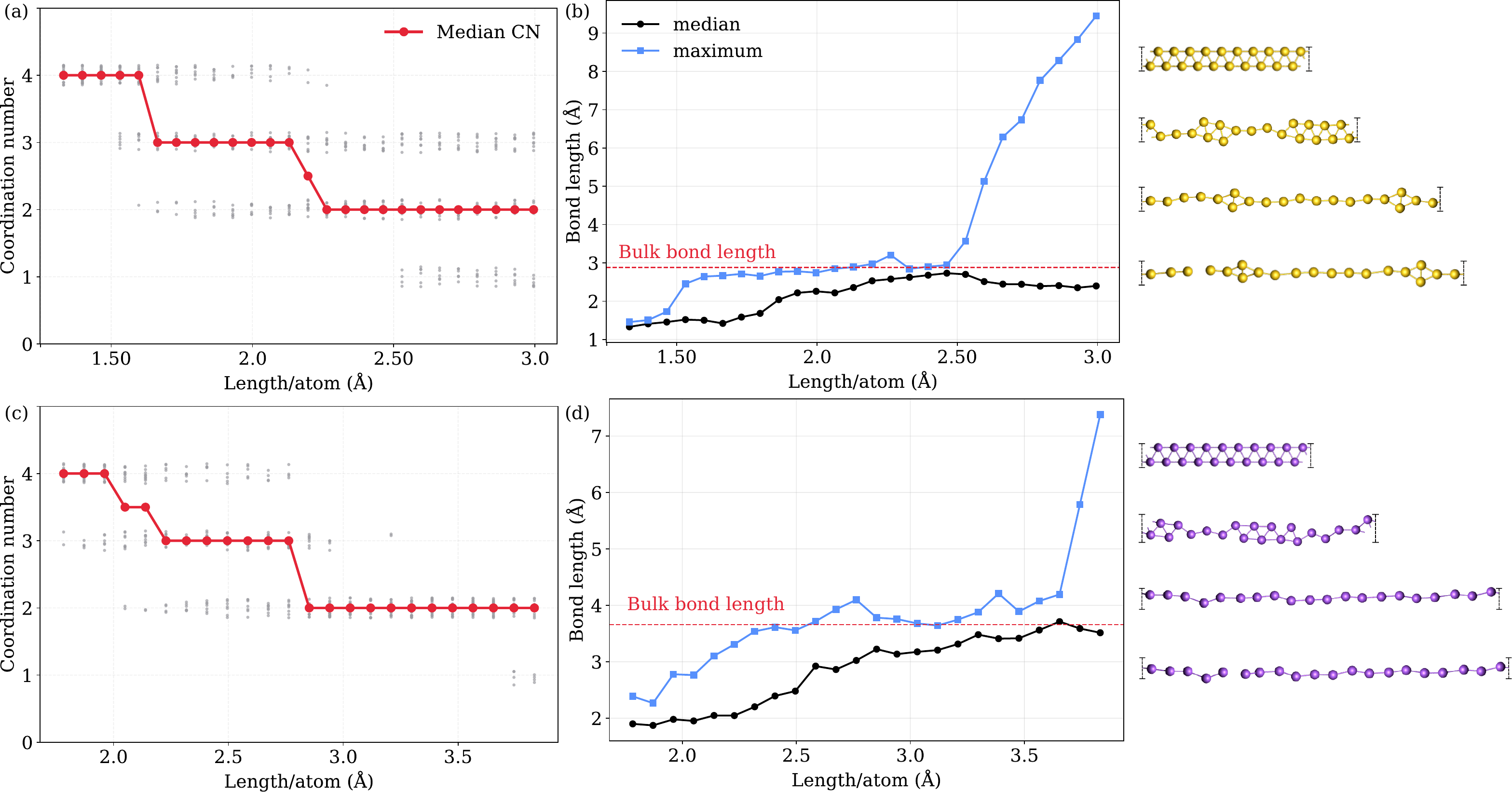}
\caption{\label{fig:pull}
 Coordination numbers, median, and maximum bond lengths in extended Au (top) and Na (bottom) chains during pulling MD simulations. The insets show snapshots of the pulling at the instants: initial supercell, at $\approx40$~\% of pulling, at $\approx90$~\% of pulling, and after breaking the first bond.}
\end{figure*}

We quantified the stability via root-mean-square displacement (RMSD) relative to the initial frame. Since chains often deformed and twisted during MD while still remained stable (see Supplementary movies), we used the Kabsch algorithm~\cite{Kabsch1976, Kabsch1978} to get an RMSD that better reflects these changes. To this end, we defined $\{\mathbf{R}_i^{(0)}\}_{i=1}^{N}$ as the reference atomic coordinates and $\{\mathbf{R}_i(t)\}_{i=1}^{N}$ as atomic coordinates from instantaneous frames. Both configurations were first translated to their geometric centers, and the instantaneous frame was then rigidly aligned to the reference using the Kabsch algorithm, which determines the optimal $3\times3$ rotation matrix $\mathbf{Q}$ that minimizes the squared distance between the two sets of atomic coordinates,
\begin{equation}
\sum_{i=1}^{N}\left|
\mathbf{R}_i^{(0,c)}-\mathbf{Q}\mathbf{R}_i^{(c)}(t)
\right|^2.
\end{equation}
Here $\mathbf{R}_i^{(0,c)}$ and $\mathbf{R}_i^{(c)}(t)$ are the centered coordinates of the reference and instantaneous frames. The aligned coordinates are then
\begin{equation}
\widetilde{\mathbf{R}}_i(t)=\mathbf{Q}\mathbf{R}_i^{(c)}(t)+\bar{\mathbf{R}}^{(0)},
\end{equation}
where $\bar{\mathbf{R}}^{(0)}$ is the geometric center of the reference frame. The RMSD is calculated as
\begin{equation}
\mathrm{RMSD}(t)=
\left[
\frac{1}{N}\sum_{i=1}^{N}
\left|\widetilde{\mathbf{R}}_i(t)-\mathbf{R}_i^{(0)}\right|^2
\right]^{1/2}
\end{equation}
characterizes the dynamical stability of the chain; values above $\sim 1$~\AA~indicate structural instability.

Based on this RMSD analysis, the distinction between stable and unstable chains becomes clear. On one hand, $14$ chains were unstable at $100$~K: Li, Na, K, Cs, Sc, Y, Ti, Zr, Hf, V, Ta, Re, Cd, and Hg (Fig.~\ref{fig:md_rmsd}b). It may well be that some chains were intrinsically unstable to begin with (dynamically unstable also at $0$~K), while for other chains the temperature may have been simply too high. On the other hand, for $26$ chains the RMSDs remained around or well below $0.5$~\AA, indicating dynamical stability. Stable chains include all alkali earth metals, all late transition metals, and post-transition metals above group $13$. These findings are further supported by radial distribution functions (RDFs) of the MD runs (Figs.~S6-S9).

We have also calculated the average bond lengths of the stable 1D chains extracted from molecular dynamics simulations (Fig.~S10). The results obtained from these supercells are well consistent with those from the ground-state unit cell, validating the two-atom cell calculations.

Finally, we performed MD simulations to pull the Au and Na chains until they broke. Using a supercell $10$ times the unit cell, we started from the chain's ground-state geometry, increased the supercell size in $5\%$ increments along the chain length with scaled atomic positions, and evolved the system for $1$~ps at each fixed length. The aim was not to simulate a continuous, dynamic chain pulling process as such, but rather to sample chain behavior at a given (mean) linear atomic density (see Supplementary movies). This simulation still enabled us to investigate the finite-temperature straightening and breaking-up dynamics of Na and Au chains.

The straightening dynamics of the two chains are different. In the Au chain, the median coordination number decreases from $4$ to $2$ (Fig.~\ref{fig:pull}a), while the maximum bond length increases rapidly upon pulling, resulting in alternating linear and tetramer-like segments (Fig.~\ref{fig:pull}b). This behavior can be understood in terms of the two local minima in the energy curve of two-atom cells (Fig.~\ref{fig:Emin_b12}a). In the Na chain, the median coordination number reduces similarly (Fig.~\ref{fig:pull}c), but the maximum bond length increases more gradually (Fig.~\ref{fig:pull}d). Also, this behavior can be understood in terms of the energy curve of a two-atom cell, which has only one minimum (Fig.~S2); the buckled geometry is more persistent, and straightening happens more uniformly. On the basis of Fig.~S2, we can then anticipate that---in addition to Au---Be, Ca, Cd, Ir, Pb, Pt, and Os chains straighten by showing more clear bifurcation to linear and tetramer segments.

\section*{Conclusions}
In conclusion, we studied $40$ non-magnetic 1D chains using density-functional theory simulations. We investigated their energetic stability, buckling, elasticity, and electronic structures. Our results revealed how extreme dimensional reduction alters metallic physics. We found that $37$ elemental chains remain metallic, and the strong quantum confinement in the 1D limit creates a metal-to-semiconductor transition in Cd, Hg, and Sr.
Structurally, we found that the perfect linear chain is rarely the ground state. For transition metals with partially filled $d$-shells, Peierls instability drives significant distortion or buckling. Transition metals retain a high fraction of their bulk cohesive energy due to the directional nature of $d$-orbitals, which adapt well to low-coordination geometries. In contrast, alkali and alkaline-earth metals, which rely on isotropic $s$-bonding and high coordination numbers, suffer a drastic loss of cohesion in 1D.

We further assessed the stability of longer 1D chains by molecular dynamics simulations. Out of the $40$ chains, $26$ maintained their structural integrity at $100$~K. Using Au and Na chains as examples, we further found that the straightening dynamics of chains can be understood in terms of minimal-cell-energy curves.

With experimental capabilities rapidly approaching the 1D chain limit, also mapping both structural stability and electronic properties trends in these free-standing models provides a foundational roadmap for identifying which elements remain conductive interconnects (e.g., Mo, W) and which evolve into optically active semiconductors (e.g., Sr) at the 1D limit of miniaturization.
While our work establishes the intrinsic physics of isolated chains, real-world realization would involve substrate support. We believe that these baseline predictions will motivate future studies investigating how substrate interactions further tune these 1D properties.

\section*{Acknowledgments}
We acknowledge the Jane and Aatos Erkko Foundation for funding (project EcoMet). We also thank the Finnish Grid and Cloud Infrastructure (FGCI) and CSC—IT Center for Science for computational resources. 

\section*{Conflict of Interest}

The authors declare no conflict of interest.

\section*{Data Availability Statement}

All data supporting the findings of this study are included within the manuscript or its Supporting Information. 
The supplementary molecular dynamics movies are available at \href{https://doi.org/10.6084/m9.figshare.32799702}{https://doi.org/10.6084/m9.figshare.32799702}.\\
Additional raw data and code used during the current study are available from the corresponding author on reasonable request.

\bibliography{refs}

@article{gpaw1,
  title = {Real-space grid implementation of the projector augmented wave method},
  author = {Mortensen, J. J. and Hansen, L. B. and Jacobsen, K. W.},
  journal = {Phys. Rev. B},
  volume = {71},
  issue = {3},
  pages = {035109},
  numpages = {11},
  year = {2005},
  month = {Jan},
  publisher = {American Physical Society},
  doi = {10.1103/PhysRevB.71.035109}
}

@article{pw,
doi = {10.1088/0953-8984/22/25/253202},
year = {2010},
month = {jun},
volume = {22},
number = {25},
pages = {253202},
author = {Enkovaara, J and Rostgaard, C and Mortensen, J J and Chen, J and Dułak, M and Ferrighi, L and Gavnholt, J and Glinsvad, C and Haikola, V and Hansen, H A and Kristoffersen, H H and Kuisma, M and Larsen, A H and Lehtovaara, L and Ljungberg, M and Lopez-Acevedo, O and Moses, P G and Ojanen, J and Olsen, T and Petzold, V and Romero, N A and Stausholm-Møller, J and Strange, M and Tritsaris, G A and Vanin, M and Walter, M and Hammer, B and Häkkinen, H and Madsen, G K H and Nieminen, R M and Nørskov, J K and Puska, M and Rantala, T T and Schiøtz, J and Thygesen, K S and Jacobsen, K W},
title = {Electronic structure calculations with GPAW: a real-space implementation of the projector augmented-wave method},
journal = {J. Phys.:Condens. Matter.}
}

@article{gpaw3,
    author = {Mortensen, Jens Jørgen and Larsen, Ask Hjorth and Kuisma, Mikael and Ivanov, Aleksei V. and Taghizadeh, Alireza and Peterson, Andrew and Haldar, Anubhab and Dohn, Asmus Ougaard and Sch\"afer, Christian and Jónsson, Elvar \"Orn and Hermes, Eric D. and Nilsson, Fredrik Andreas and Kastlunger, Georg and Levi, Gianluca and Jónsson, Hannes and Häkkinen, Hannu and Fojt, Jakub and Kangsabanik, Jiban and Sødequist, Joachim and Lehtomäki, Jouko and Heske, Julian and Enkovaara, Jussi and Winther, Kirsten Trøstrup and Dulak, Marcin and Melander, Marko M. and Ovesen, Martin and Louhivuori, Martti and Walter, Michael and Gjerding, Morten and Lopez-Acevedo, Olga and Erhart, Paul and Warmbier, Robert and Würdemann, Rolf and Kaappa, Sami and Latini, Simone and Boland, Tara Maria and Bligaard, Thomas and Skovhus, Thorbjørn and Susi, Toma and Maxson, Tristan and Rossi, Tuomas and Chen, Xi and Schmerwitz, Yorick Leonard A. and Schiøtz, Jakob and Olsen, Thomas and Jacobsen, Karsten Wedel and Thygesen, Kristian Sommer},
    title = {GPAW: An open Python package for electronic structure calculations},
    journal = {J. Chem. Phys.},
    volume = {160},
    number = {9},
    pages = {092503},
    year = {2024},
    month = {03},
    issn = {0021-9606},
    doi = {10.1063/5.0182685}
}

@article{libxc,
title = {Recent developments in libxc — A comprehensive library of functionals for density functional theory},
journal = {SoftwareX},
volume = {7},
pages = {1-5},
year = {2018},
issn = {2352-7110},
doi = {https://doi.org/10.1016/j.softx.2017.11.002},
author = {Susi Lehtola and Conrad Steigemann and Micael J.T. Oliveira and Miguel A.L. Marques},
}

@article{ase-paper,
  author={Ask Hjorth Larsen and Jens Jørgen Mortensen and Jakob Blomqvist and Ivano E Castelli and Rune Christensen and Marcin
Dułak and Jesper Friis and Michael N Groves and Bjørk Hammer and Cory Hargus and Eric D Hermes and Paul C Jennings and Peter
Bjerre Jensen and James Kermode and John R Kitchin and Esben Leonhard Kolsbjerg and Joseph Kubal and Kristen
Kaasbjerg and Steen Lysgaard and Jón Bergmann Maronsson and Tristan Maxson and Thomas Olsen and Lars Pastewka and Andrew
Peterson and Carsten Rostgaard and Jakob Schiøtz and Ole Schütt and Mikkel Strange and Kristian S Thygesen and Tejs
Vegge and Lasse Vilhelmsen and Michael Walter and Zhenhua Zeng and Karsten W Jacobsen},
  title={The atomic simulation environment—a Python library for working with atoms},
  journal={J. Phys.:Condens. Matter.},
  volume={29},
  number={27},
  pages={273002},
  year={2017}
}

@article{pbe,
  title = {Generalized Gradient Approximation Made Simple},
  author = {Perdew, John P. and Burke, Kieron and Ernzerhof, Matthias},
  journal = {Phys. Rev. Lett.},
  volume = {77},
  issue = {18},
  pages = {3865--3868},
  numpages = {0},
  year = {1996},
  month = {Oct},
  publisher = {American Physical Society},
  doi = {10.1103/PhysRevLett.77.3865}
}

@article{GGA,
  title = {Generalized Gradient Approximation Made Simple [Phys. Rev. Lett. 77, 3865 (1996)]},
  author = {Perdew, John P. and Burke, Kieron and Ernzerhof, Matthias},
  journal = {Phys. Rev. Lett.},
  volume = {78},
  issue = {7},
  pages = {1396--1396},
  numpages = {0},
  year = {1997},
  month = {Feb},
  publisher = {American Physical Society},
  doi = {10.1103/PhysRevLett.78.1396}
}

@article{PhysRevB.13.5188,
  title = {Special points for Brillouin-zone integrations},
  author = {Monkhorst, Hendrik J. and Pack, James D.},
  journal = {Phys. Rev. B},
  volume = {13},
  issue = {12},
  pages = {5188--5192},
  numpages = {0},
  year = {1976},
  month = {Jun},
  publisher = {American Physical Society},
  doi = {10.1103/PhysRevB.13.5188}
}

@article{PhysRevB.16.1748,
  title = {"Special points for Brillouin-zone integrations"---a reply},
  author = {Pack, James D. and Monkhorst, Hendrik J.},
  journal = {Phys. Rev. B},
  volume = {16},
  issue = {4},
  pages = {1748--1749},
  numpages = {0},
  year = {1977},
  month = {Aug},
  publisher = {American Physical Society},
  doi = {10.1103/PhysRevB.16.1748}
}

@article{d3-1,
    author = {Grimme, Stefan and Antony, Jens and Ehrlich, Stephan and Krieg, Helge},
    title = {A consistent and accurate ab initio parametrization of density functional dispersion correction (DFT-D) for the 94 elements H-Pu},
    journal = {The Journal of Chemical Physics},
    volume = {132},
    number = {15},
    pages = {154104},
    year = {2010},
    issn = {0021-9606},
    doi = {https://doi.org/10.1063/1.3382344}
}

@article{d3-2,
author = {Grimme, Stefan and Ehrlich, Stephan and Goerigk, Lars},
title = {Effect of the damping function in dispersion corrected density functional theory},
journal = {Journal of Computational Chemistry},
volume = {32},
number = {7},
pages = {1456-1465},
doi = {https://doi.org/10.1002/jcc.21759},
year = {2011}
}

@article{d3-3,
author = {Smith, Daniel G. A. and Burns, Lori A. and Patkowski, Konrad and Sherrill, C. David},
title = {Revised Damping Parameters for the D3 Dispersion Correction to Density Functional Theory},
journal = {The Journal of Physical Chemistry Letters},
volume = {7},
number = {12},
pages = {2197-2203},
year = {2016},
doi = {https://doi.org/10.1021/acs.jpclett.6b00780} 
}

@article{atlas_2018,
  title={Atlas for the properties of elemental two-dimensional metals},
  author={Nevalaita, Janne and Koskinen, Pekka},
  journal={Phys. Rev. B},
  volume={97},
  number={3},
  pages={035411},
  year={2018},
  publisher={APS}
}

@article{SCHMIDT_1997,
title = {One-dimensional chains of metal atoms},
journal = {Solid State Communications},
volume = {104},
number = {7},
pages = {413-417},
year = {1997},
issn = {0038-1098},
doi = {https://doi.org/10.1016/S0038-1098(97)00345-1},
author = {Karla Schmidt and Michael Springborg}
}

@article{1D_gold_2004,
  title = {Chains of gold atoms with tailored electronic states},
  author = {Crain, J. N. and McChesney, J. L. and Zheng, Fan and Gallagher, M. C. and Snijders, P. C. and Bissen, M. and Gundelach, C. and Erwin, S. C. and Himpsel, F. J.},
  journal = {Phys. Rev. B},
  volume = {69},
  issue = {12},
  pages = {125401},
  numpages = {10},
  year = {2004},
  month = {Mar},
  publisher = {American Physical Society},
  doi = {10.1103/PhysRevB.69.125401},
  url = {https://link.aps.org/doi/10.1103/PhysRevB.69.125401}
}

@article{MA_2022,
title = {Structure, synthesis, and properties of single-metal-atom chains},
journal = {Cell Reports Physical Science},
volume = {3},
number = {11},
pages = {101124},
year = {2022},
issn = {2666-3864},
doi = {https://doi.org/10.1016/j.xcrp.2022.101124},
author = {Mingyu Ma and Shasha Guo and Xiaoru Sang and Caitian Gao and Zheng Liu and Yongmin He}
}

@article{Gambardella_2002,
  title = {Ferromagnetism in one-dimensional monatomic metal chains},
  volume = {416},
  ISSN = {1476-4687},
  DOI = {http://doi.org/10.1038/416301a},
  number = {6878},
  journal = {Nature},
  publisher = {Springer Science and Business Media LLC},
  author = {Gambardella,  P. and Dallmeyer,  A. and Maiti,  K. and Malagoli,  M. C. and Eberhardt,  W. and Kern,  K. and Carbone,  C.},
  year = {2002},
  pages = {301–304}
}

@article{BALANDIN_2022,
title = {One-dimensional van der Waals quantum materials},
journal = {Materials Today},
volume = {55},
pages = {74-91},
year = {2022},
issn = {1369-7021},
doi = {https://doi.org/10.1016/j.mattod.2022.03.015},
author = {Alexander A. Balandin and Fariborz Kargar and Tina T. Salguero and Roger K. Lake}
}

@article{Yanbing_2023,
author = {Zhu, Yanbing and Antoniuk, Evan R. and Wright, Dylan and Kargar, Fariborz and Sesing, Nicholas and Sendek, Austin D. and Salguero, Tina T. and Bartels, Ludwig and Balandin, Alexander A. and Reed, Evan J. and da Jornada, Felipe H.},
title = {Machine-Learning-Driven Expansion of the 1D van der Waals Materials Space},
journal = {J. Phys. Chem. C},
volume = {127},
number = {44},
pages = {21675-21683},
year = {2023},
doi = {10.1021/acs.jpcc.3c03882}
}

@article{Hadeel_2023,
  title = {Hundreds of new, stable, one-dimensional materials from a generative machine learning model},
  author = {Moustafa, Hadeel and Lyngby, Peder Meisner and Mortensen, Jens J\o{}rgen and Thygesen, Kristian S. and Jacobsen, Karsten W.},
  journal = {Phys. Rev. Mater.},
  volume = {7},
  issue = {1},
  pages = {014007},
  numpages = {10},
  year = {2023},
  publisher = {American Physical Society},
  doi = {10.1103/PhysRevMaterials.7.014007},
  url = {https://link.aps.org/doi/10.1103/PhysRevMaterials.7.014007}
}

@article{Khazaei_2023,
author = {Khazaei, Mohammad and Bae, Soungmin and Khaledialidusti, Rasoul and Ranjbar, Ahmad and Komsa, Hannu-Pekka and Khazaei, Somayeh and Bagheri, Mohammad and Wang, Vei and Mochizuki, Yasuhide and Kawamura, Mitsuaki and Cuniberti, Gianaurelio and Allaei, S. Mehdi Vaez and Ohno, Kaoru and Hosono, Hideo and Raebiger, Hannes},
title = {Superlattice MAX Phases with A-Layers Reconstructed into 0D-Clusters, 1D-Chains, and 2D-Lattices},
journal = {J. Phys. Chem. C},
volume = {127},
number = {30},
pages = {14906-14913},
year = {2023},
doi = {10.1021/acs.jpcc.3c02233}
}

@article{Teeter_2024,
author = {Teeter, Jordan and Kim, Na Yeon and Debnath, Topojit and Sesing, Nicholas and Geremew, Tekwam and Wright, Dylan and Chi, Miaofang and Stieg, Adam Z. and Miao, Jianwei and Lake, Roger K. and Salguero, Tina and Balandin, Alexander A.},
title = {Achieving the 1D Atomic Chain Limit in Van der Waals Crystals},
journal = {Adv. Mater.},
volume = {36},
number = {48},
pages = {2409898},
doi = {https://doi.org/10.1002/adma.202409898},
year = {2024}
}

@article{FCDimen,
author = {Bagheri, Mohammad and Berger, Ethan and Komsa, Hannu-Pekka},
title = {Identification of Material Dimensionality Based on Force Constant Analysis},
journal = {J. Phys. Chem. Lett.},
volume = {14},
number = {35},
pages = {7840-7847},
year = {2023},
doi = {10.1021/acs.jpclett.3c01635}
}

@article{MLFCDimen,
      title={Massive Discovery of Low-Dimensional Materials from Universal Computational Strategy}, 
      author={Mohammad Bagheri and Ethan Berger and Hannu-Pekka Komsa and Pekka Koskinen},
    journal = {Chemistry of Materials},
     volume = {38},
     number = {5},
     pages = {2395–-2402},
     year = {2026},
     doi = {https://doi.org/10.1021/acs.chemmater.5c03151}
}

@article{Crain_2004,
  title = {Chains of gold atoms with tailored electronic states},
  author = {Crain, J. N. and McChesney, J. L. and Zheng, Fan and Gallagher, M. C. and Snijders, P. C. and Bissen, M. and Gundelach, C. and Erwin, S. C. and Himpsel, F. J.},
  journal = {Phys. Rev. B},
  volume = {69},
  issue = {12},
  pages = {125401},
  numpages = {10},
  year = {2004},
  publisher = {American Physical Society},
  doi = {10.1103/PhysRevB.69.125401},
}

@article{Erwin_2013,
  title = {Silicon spin chains at finite temperature: Dynamics of Si(553)-Au},
  author = {Erwin, Steven C. and Snijders, Paul C.},
  journal = {Phys. Rev. B},
  volume = {87},
  issue = {23},
  pages = {235316},
  numpages = {8},
  year = {2013},
  publisher = {American Physical Society},
  doi = {10.1103/PhysRevB.87.235316}
}

@article{Polei_2013,
  title = {Structural Transition in Atomic Chains Driven by Transient Doping},
  author = {Polei, S. and Snijders, P. C. and Erwin, S. C. and Himpsel, F. J. and Meiwes-Broer, K-H. and Barke, I.},
  journal = {Phys. Rev. Lett.},
  volume = {111},
  issue = {15},
  pages = {156801},
  numpages = {5},
  year = {2013},
  publisher = {American Physical Society},
  doi = {10.1103/PhysRevLett.111.156801},
}

@article{Anasori2017,
  title = {2D metal carbides and nitrides (MXenes) for energy storage},
  volume = {2},
  ISSN = {2058-8437},
  DOI = {10.1038/natrevmats.2016.98},
  number = {2},
  journal = {Nat. Rev. Mater.},
  publisher = {Springer Science and Business Media LLC},
  author = {Anasori,  Babak and Lukatskaya,  Maria R. and Gogotsi,  Yury},
  year = {2017}
}

@article{Guan_2017,
author = {Guan, Guijian and Xia, Jing and Liu, Shuhua and Cheng, Yuan and Bai, Shiqiang and Tee, Si Yin and Zhang, Yong-Wei and Han, Ming-Yong},
title = {Electrostatic-Driven Exfoliation and Hybridization of 2D Nanomaterials},
journal = {Adv. Mater.},
volume = {29},
number = {32},
pages = {1700326},
doi = {https://doi.org/10.1002/adma.201700326},
year = {2017}
}

@article{Balan_2018,
author = {Puthirath Balan, Aravind and Radhakrishnan, Sruthi and Kumar, Ritesh and Neupane, Ram and Sinha, Shyam Kanta and Deng, Liangzi and de los Reyes, Carlos A. and Apte, Amey and Rao, B. Manmadha and Paulose, Maggie and Vajtai, Robert and Chu, Ching Wu and Costin, Gelu and Martí, Angel A. and Varghese, Oomman K. and Singh, Abhishek K. and Tiwary, Chandra Sekhar and Anantharaman, Maliemadom R. and Ajayan, Pulickel M.},
title = {A Non-van der Waals Two-Dimensional Material from Natural Titanium Mineral Ore Ilmenite},
journal = {Chem. Mater.},
volume = {30},
number = {17},
pages = {5923-5931},
year = {2018},
doi = {10.1021/acs.chemmater.8b01935}
}

@article{PuthirathBalan2018,
  title = {Exfoliation of a non-van der Waals material from iron ore hematite},
  volume = {13},
  ISSN = {1748-3395},
  DOI = {10.1038/s41565-018-0134-y},
  number = {7},
  journal = {Nat. Nanotechnol.},
  publisher = {Springer Science and Business Media LLC},
  author = {Puthirath Balan,  Aravind and Radhakrishnan,  Sruthi and Woellner,  Cristiano F. and Sinha,  Shyam K. and Deng,  Liangzi and Reyes,  Carlos de los and Rao,  Banki Manmadha and Paulose,  Maggie and Neupane,  Ram and Apte,  Amey and Kochat,  Vidya and Vajtai,  Robert and Harutyunyan,  Avetik R. and Chu,  Ching-Wu and Costin,  Gelu and Galvao,  Douglas S. and Martí,  Angel A. and van Aken,  Peter A. and Varghese,  Oomman K. and Tiwary,  Chandra Sekhar and Malie Madom Ramaswamy Iyer,  Anantharaman and Ajayan,  Pulickel M.},
  year = {2018},
  pages = {602–609}
}

@article{Jiang2022,
  title = {Mechanical cleavage of non-van der Waals structures towards two-dimensional crystals},
  volume = {2},
  ISSN = {2731-0582},
  DOI = {10.1038/s44160-022-00182-6},
  number = {1},
  journal = {Nat. Synth.},
  publisher = {Springer Science and Business Media LLC},
  author = {Jiang,  Kun and Ji,  Jinpeng and Gong,  Wenbin and Ding,  Ling and Li,  Jibiao and Li,  Pengfei and Li,  Baowen and Geng,  Fengxia},
  year = {2022},
  pages = {58–66}
}

@article{Kaur_2022,
author = {Kaur, Harneet and Coleman, Jonathan N.},
title = {Liquid-Phase Exfoliation of Nonlayered Non-Van-Der-Waals Crystals into Nanoplatelets},
journal = {Adv. Mater.},
volume = {34},
number = {35},
pages = {2202164},
doi = {https://doi.org/10.1002/adma.202202164},
year = {2022}
}

@article{Radisavljevic2011,
  doi = {10.1038/nnano.2010.279},
  year = {2011},
  publisher = {Springer Science and Business Media {LLC}},
  volume = {6},
  number = {3},
  pages = {147-150},
  author = {B. Radisavljevic and A. Radenovic and J. Brivio and V. Giacometti and A. Kis},
  title = {Single-Layer {M}o{S}$_2$ Transistors},
  journal = {Nat. Nanotechnol.}
}

@article{Manzeli2017,
  doi = {10.1038/natrevmats.2017.33},
  year = {2017},
  publisher = {Springer Science and Business Media {LLC}},
  volume = {2},
  pages = {17033},
  author = {Sajedeh Manzeli and Dmitry Ovchinnikov and Diego Pasquier and Oleg V. Yazyev and Andras Kis},
  title = {2{D} Transition Metal Dichalcogenides},
  journal = {Nat. Rev. Mater.}
}

@article{Xiang20_Sci,
  author    = {Xiang, Rong and Inoue, Taiki and Zheng, Yongjia and Kumamoto, Akihito and Qian, Yang and Sato, Yuta and Liu, Ming and Tang, Daiming and Gokhale, Devashish and Guo, Jia and Hisama, Kaoru and Yotsumoto, Satoshi and Ogamoto, Tatsuro and Arai, Hayato and Kobayashi, Yu and Zhang, Hao and Hou, Bo and Anisimov, Anton and Maruyama, Mina and Miyata, Yasumitsu and Okada, Susumu and Chiashi, Shohei and Li, Yan and Kong, Jing and Kauppinen, Esko I. and Ikuhara, Yuichi and Suenaga, Kazu and Maruyama, Shigeo},
  title     = {One-Dimensional Van der {W}aals Heterostructures},
  journal   = {Science},
  year      = {2020},
  volume    = {367},
  number    = {6477},
  pages     = {537-542},
  issn      = {0036-8075},
  doi       = {10.1126/science.aaz2570},
  publisher = {American Association for the Advancement of Science}
}

@article{Han19_NComm,
  author  = {Han, Wei and Huang, Pu and Li, Liang and Wang, Fakun and Luo, Peng and Liu, Kailang and Zhou, Xing and Li, Huiqiao and Zhang, Xiuwen and Cui, Yi and Zhai, Tianyou},
  title   = {Two-Dimensional Inorganic Molecular Crystals},
  journal = {Nat. Commun.},
  year    = {2019},
  volume  = {10},
  number  = {1},
  pages   = {4728},
  doi     = {10.1038/s41467-019-12569-9},
}

@article{Du21_NComm,
	author={Du, Luojun and Zhao, Yanchong and Wu, Linlu and Hu, Xuerong and Yao, Lide and Wang, Yadong and Bai, Xueyin and Dai, Yunyun and Qiao, Jingsi and Uddin, Md Gius and Li, Xiaomei and Lahtinen, Jouko and Bai, Xuedong and Zhang, Guangyu and Ji, Wei and Sun, Zhipei},
	title={Giant Anisotropic Photonics in the 1{D} Van der Waals Semiconductor Fibrous Red Phosphorus},
    journal={Nat. Commun.},
	pages={4822},
	year={2021},
	doi={10.1038/s41467-021-25104-6},
	volume={12},
	number={1},
}

@article{Wei21_AM,
author = {Wei, Nan and Tian, Ying and Liao, Yongping and Komatsu, Natsumi and Gao, Weilu and Lyuleeva-Husemann, Alina and Zhang, Qiang and Hussain, Aqeel and Ding, Er-Xiong and Yao, Fengrui and Halme, Janne and Liu, Kaihui and Kono, Junichiro and Jiang, Hua and Kauppinen, Esko I.},
title = {Colors of Single-wall Carbon Nanotubes},
journal = {Adv. Mater.},
volume = {33},
number = {8},
pages = {2006395},
doi = {10.1002/adma.202006395},
year = {2021}
}

@article{Turunen22_NRevPhys,
  author  = {Turunen, Mikko and Brotons-Gisbert, Mauro and Dai, Yunyun and Wang, Yadong and Scerri, Eleanor and Bonato, Cristian and Jöns, Klaus D. and Sun, Zhipei and Gerardot, Brian D.},
  title   = {Quantum Photonics with Layered 2{D} Materials},
  journal = {Nat. Rev. Phys.},
  year    = {2022},
  volume  = {4},
  pages   = {219},
  doi     = {10.1038/s42254-021-00408-0}
}

@article{Gibertini2019,
  doi = {https://doi.org/10.1038/s41565-019-0438-6},
  year = {2019},
  publisher = {Springer Science and Business Media {LLC}},
  volume = {14},
  number = {5},
  pages = {408-419},
  author = {M. Gibertini and M. Koperski and A. F. Morpurgo and K. S. Novoselov},
  title = {Magnetic 2{D} Materials and Heterostructures},
  journal = {Nat. Nanotechnol.}
}

@article{Klein23_ACSNano,
author = {Klein, Julian and Pingault, Benjamin and Florian, Matthias and Heißenbüttel, Marie-Christin and Steinhoff, Alexander and Song, Zhigang and Torres, Kierstin and Dirnberger, Florian and Curtis, Jonathan B. and Weile, Mads and Penn, Aubrey and Deilmann, Thorsten and Dana, Rami and Bushati, Rezlind and Quan, Jiamin and Luxa, Jan and Sofer, Zdeněk and Alù, Andrea and Menon, Vinod M. and Wurstbauer, Ursula and Rohlfing, Michael and Narang, Prineha and Lončar, Marko and Ross, Frances M.},
title = {The Bulk Van der Waals Layered Magnet {C}r{SB}r is A Quasi-1{D} Material},
journal = {ACS Nano},
volume = {17},
number = {6},
pages = {5316-5328},
year = {2023},
doi = {10.1021/acsnano.2c07316}
}

@article{Kim_2018,
author = {Kim, Seon Joon and Koh, Hyeong-Jun and Ren, Chang E. and Kwon, Ohmin and Maleski, Kathleen and Cho, Soo-Yeon and Anasori, Babak and Kim, Choong-Ki and Choi, Yang-Kyu and Kim, Jihan and Gogotsi, Yury and Jung, Hee-Tae},
title = {Metallic {T}i$_3${C}$_2${T}$_x$ {MX}ene Gas Sensors with Ultrahigh Signal-to-Noise Ratio},
journal = {ACS Nano},
volume = {12},
number = {2},
pages = {986-993},
year = {2018},
doi = {10.1021/acsnano.7b07460}
}

@article{mx_sens,
author = {Hosseini-Shokouh, Seyed Hossein and Zhou, Jin and Berger, Ethan and Lv, Zhong-Peng and Hong, Xiaodan and Virtanen, Vesa and Kordas, Krisztian and Komsa, Hannu-Pekka},
title = {Highly Selective {H}$_2${S} Gas Sensor Based on {T}i$_3${C}$_2${T}$_x$ {MX}ene–Organic Composites},
journal = {ACS Appl. Mater. Interfaces},
volume = {15},
number = {5},
pages = {7063-7073},
year = {2023},
doi = {10.1021/acsami.2c19883}
}

@article{Zhou_2021,
author = {Zhou, Jin and Bagheri, Mohammad and Järvinen, Topias and Pravda Bartus, Cora and Kukovecz, Akos and Komsa, Hannu-Pekka and Kordas, Krisztian},
title = {{C}$_{60}${B}r$_{24}$/{SWCNT}: A Highly Sensitive Medium to Detect H$_2$S via Inhomogeneous Carrier Doping},
journal = {ACS Appl. Mater. Interfaces},
volume = {13},
number = {49},
pages = {59067-59075},
year = {2021},
doi = {10.1021/acsami.1c16807}
}

@article{Deng2016,
  title = {Catalysis with Two-Dimensional Materials and Their Heterostructures},
  volume = {11},
  ISSN = {1748-3395},
  DOI = {10.1038/nnano.2015.340},
  number = {3},
  journal = {Nat. Nanotechnol.},
  publisher = {Springer Science and Business Media LLC},
  author = {Deng,  Dehui and Novoselov,  K. S. and Fu,  Qiang and Zheng,  Nanfeng and Tian,  Zhongqun and Bao,  Xinhe},
  year = {2016},
  pages = {218–230}
}

@article{D4NR00147H,
author ={Cho, Yun Seong and Kang, Joohoon},
title  ={Two-Dimensional Materials as Catalysts{,} Interfaces{,} and Electrodes for An Efficient Hydrogen Evolution Reaction},
journal  ={Nanoscale},
year  ={2024},
volume  ={16},
issue  ={8},
pages  ={3936-3950},
publisher  ={The Royal Society of Chemistry},
doi  ={10.1039/D4NR00147H}
}

@article{Zhang_2025,
author = {Zhang, Mo and Wang, Zifeng and Bo, Xin and Huang, Rui and Deng, Dehui},
title = {Two-Dimensional Catalysts: From Model to Reality},
journal = {Angew. Chem., Int. Ed.},
volume = {64},
number = {5},
pages = {e202419661},
doi = {https://doi.org/10.1002/anie.202419661},
year = {2025}
}

@article{TAO2020,
title = {Two-dimensional materials for energy conversion and storage},
journal = {Progress in Materials Science},
volume = {111},
pages = {100637},
year = {2020},
issn = {0079-6425},
doi = {https://doi.org/10.1016/j.pmatsci.2020.100637},
author = {Hengcong Tao and Qun Fan and Tao Ma and Shizhen Liu and Henry Gysling and John Texter and Fen Guo and Zhenyu Sun}
}

@article{Qi2024,
author = {Qi, Junlei and Bao, Kai and Wang, Wenbin and Wu, Jingkun and Wang, Lingzhi and Ma, Cong and Wu, Zongxiao and He, Qiyuan},
title = {Emerging Two-Dimensional Materials for Proton-Based Energy Storage},
journal = {ACS Nano},
volume = {18},
number = {38},
pages = {25910-25929},
year = {2024},
doi = {10.1021/acsnano.4c06737}
}

@article{gentle,
author ={Abidi, Kameyab Raza and Koskinen, Pekka},
title  ={Gentle tension stabilizes atomically thin metallenes},
journal  ={Nanoscale},
year  ={2024},
volume  ={16},
issue  ={42},
pages  ={19649-19655},
publisher  ={The Royal Society of Chemistry},
doi  ={10.1039/D4NR03266G}
}

@article{interface,
author ="Bagheri, Mohammad and Koskinen, Pekka",
title  ="Lateral graphene-metallene interfaces at the nanoscale",
journal  ="Nanoscale",
year  ="2026",
volume  ="18",
issue  ="1",
pages  ="188-196",
publisher  ="The Royal Society of Chemistry",
doi  ="10.1039/D5NR02770E"
}

@article{HSE06,
    author = {Heyd, Jochen and Scuseria, Gustavo E. and Ernzerhof, Matthias},
    title = {Hybrid functionals based on a screened Coulomb potential},
    journal = {The Journal of Chemical Physics},
    volume = {118},
    number = {18},
    pages = {8207-8215},
    year = {2003},
    doi = {10.1063/1.1564060},
}

@article{hg_2006,
  title = {Lattice structure of mercury: Influence of electronic correlation},
  author = {Gaston, Nicola and Paulus, Beate and Rosciszewski, Krzysztof and Schwerdtfeger, Peter and Stoll, Hermann},
  journal = {Phys. Rev. B},
  volume = {74},
  issue = {9},
  pages = {094102},
  numpages = {9},
  year = {2006},
  month = {Sep},
  publisher = {American Physical Society},
  doi = {10.1103/PhysRevB.74.094102}
}

@article{fermi_met,
      title={Lattice and Orbital-Resolved Fermiology of Metallenes}, 
      author={Kameyab Raza Abidi and Mohammad Bagheri and Pekka Koskinen},
 journal = {Phys. Rev. B},
  volume = {113},
  issue = {11},
  pages = {115117},
  numpages = {13},
  year = {2026},
  publisher = {American Physical Society},
  doi = {10.1103/lnm9-p3yl}
}

@book{peierls1955quantum,
  title={Quantum theory of solids},
  author={Peierls, Rudolf Ernst},
  year={1955},
  publisher={Oxford University Press}
}

@article{Abidi_2025,
doi = {10.1088/2516-1075/adb4b5},
url = {https://doi.org/10.1088/2516-1075/adb4b5},
year = {2025},
month = {feb},
publisher = {IOP Publishing},
volume = {7},
number = {1},
pages = {015004},
author = {Abidi, Kameyab Raza and Koskinen, Pekka},
title = {Electronic and structural properties of atomically thin metallenes},
journal = {Electronic Structure}
}

@article{D1TC00438G,
author ={Ren, Yinti and Hu, Liang and Shao, Yangfan and Hu, Yijian and Huang, Li and Shi, Xingqiang},
title  ={Magnetism of elemental two-dimensional metals},
journal  ={J. Mater. Chem. C},
year  ={2021},
volume  ={9},
issue  ={13},
pages  ={4554-4561},
publisher  ={The Royal Society of Chemistry},
doi  ={10.1039/D1TC00438G}
}

@article{abidi2022optimizing,
  title={Optimizing density-functional simulations for two-dimensional metals},
  author={Abidi, Kameyab Raza and Koskinen, Pekka},
  journal={Physical Review Materials},
  volume={6},
  number={12},
  pages={124004},
  year={2022},
  publisher={APS}
}

@article{Edge_2025,
doi = {10.1088/2053-1583/adafc1},
year = {2025},
month = {feb},
publisher = {IOP Publishing},
volume = {12},
number = {2},
pages = {025016},
author = {Abidi, Kameyab Raza and Bagheri, Mohammad and Singh, Sukhbir and Koskinen, Pekka},
title = {Atomically thin metallenes at the edge},
journal = {2D Mater.}
}

@article{Kabsch1976,
  author  = {Kabsch, Wolfgang},
  title   = {A solution for the best rotation to relate two sets of vectors},
  journal = {Acta Crystallographica Section A},
  volume  = {32},
  pages   = {922--923},
  year    = {1976},
  doi = {10.1107/S0567739476001873}
}

@article{Kabsch1978,
  author  = {Kabsch, Wolfgang},
  title   = {A discussion of the solution for the best rotation to relate two sets of vectors},
  journal = {Acta Crystallographica Section A},
  volume  = {34},
  pages   = {827--828},
  year    = {1978},
  doi = {10.1107/S0567739478001680}
}

@article{Mihalyuk_2020,
  title = {One-dimensional Rashba states in Pb atomic chains on a semiconductor surface},
  author = {Mihalyuk, A. N. and Chou, J. P. and Eremeev, S. V. and Zotov, A. V. and Saranin, A. A.},
  journal = {Phys. Rev. B},
  volume = {102},
  issue = {3},
  pages = {035442},
  numpages = {8},
  year = {2020},
  month = {Jul},
  publisher = {American Physical Society},
  doi = {10.1103/PhysRevB.102.035442}
}

@article{Sheverdyaeva_2022,
  title = {One-dimensional Rashba states with unconventional spin texture in Bi chains},
  author = {Sheverdyaeva, P. M. and Pacil\`e, D. and Topwal, D. and Manju, U. and Papagno, M. and Feyer, V. and Jugovac, M. and Zamborlini, G. and Cojocariu, I. and Tusche, C. and Tan, X. L. and Hagiwara, K. and Chen, Y.-J. and Fujii, J. and Moras, P. and Ferrari, L. and Vescovo, E. and Bihlmayer, G. and Carbone, C.},
  journal = {Phys. Rev. B},
  volume = {106},
  issue = {4},
  pages = {045108},
  numpages = {6},
  year = {2022},
  month = {Jul},
  publisher = {American Physical Society},
  doi = {10.1103/PhysRevB.106.045108}
}

@article{Mihalyuk_2024,
    author = {Mihalyuk, Alexey N. and Bondarenko, Leonid V. and Tupchaya, Alexandra Y. and Gruznev, Dimitry V. and Solovova, Nadezhda Yu. and Golyashov, Vladimir A. and Tereshchenko, Oleg E. and Okuda, Taichi and Kimura, Akio and Eremeev, Sergey V. and Zotov, Andrey V. and Saranin, Alexander A.},
    title = {Emergence of quasi-1D spin-polarized states in ultrathin Bi films on InAs(111)A for spintronics applications},
    journal = {Nanoscale},
    volume = {16},
    number = {3},
    pages = {1272-1281},
    year = {2024},
    issn = {2040-3364},
    doi = {10.1039/d3nr03830k}
}

@article{Sheverdyaeva_2024,
    author = {Sheverdyaeva, Polina M. and Bihlmayer, Gustav and Modesti, Silvio and Feyer, Vitaliy and Jugovac, Matteo and Zamborlini, Giovanni and Tusche, Christian and Chen, Ying-Jiun and Tan, Xin Liang and Hagiwara, Kenta and Petaccia, Luca and Thakur, Sangeeta and Kundu, Asish K. and Carbone, Carlo and Moras, Paolo},
    title = {Giant Rashba-splitting of one-dimensional metallic states in Bi dimer lines on InAs(100)},
    journal = {Nanoscale},
    volume = {16},
    number = {33},
    pages = {15815-15823},
    year = {2024},
    issn = {2040-3364},
    doi = {10.1039/d4nr01591f},
}

@article{PFNUR_2024,
title = {Atomic wires on substrates: Physics between one and two dimensions},
journal = {Surface Science Reports},
volume = {79},
number = {2},
pages = {100629},
year = {2024},
issn = {0167-5729},
doi = {https://doi.org/10.1016/j.surfrep.2024.100629},
author = {H. Pfnür and C. Tegenkamp and S. Sanna and E. Jeckelmann and M. {Horn-von Hoegen} and U. Bovensiepen and N. Esser and W.G. Schmidt and M. Dähne and S. Wippermann and F. Bechstedt and M. Bode and R. Claessen and R. Ernstorfer and C. Hogan and M. Ligges and A. Pucci and J. Schäfer and E. Speiser and M. Wolf and J. Wollschläger},
}

\end{document}